  \documentclass[final,3p,times,twocolumn]{elsarticle}
\usepackage{float}

\journal{Physics Letters A}

\usepackage{graphicx}
 \usepackage{epsfig}
\begin{document}
\def\ds{\displaystyle}
\begin{frontmatter}

\title{Topological phase in a non-Hermitian PT symmetric system}
\author{C. Yuce }
\address{ Physics Department, Anadolu University, Eskisehir, Turkey}
\ead{cyuce@anadolu.edu.tr} \fntext[label2]{}
\begin{abstract}
In this work, we consider a tight binding lattice with two non-Hermitian impurities. The system is described by a non-Hermitian generalization of the Aubry Andre model. We show for the first time that there exists topologically nontrivial edge states with real spectra in the $\mathcal{PT}$ symmetric region.
\end{abstract}


\end{frontmatter}


\section{Introduction}

The discovery of topological insulators in 2D and 3D has recently attracted a great deal of attention \cite{hasan} (reference therein). The appearance of the gapless edge states within the bulk gap is generally believed to be a signature of the topological insulator. Generally speaking, a topological insulator has gapless robust edge states while it's energy spectrum is gapped in the bulk. The bulk energy gap closes if a system is deformed adiabatically to a topologically nonequivalent system. Remarkably, topological phase is not restricted to two and three dimensional systems. Of special importance in the context of localization and topological phase in 1D is the Aubry and Andre (AA) model \cite{andre,10,18,A2,16,AAH,A1,A3,A4,A5,A6,A7}. A mapping of the 1D AA model to topologically-nontrivial 2D quantum Hall (QH) system was constructed and used to topologically classify 1D quasicrystals \cite{10,18}. Topological equivalence between crystal and quasicrystal band structures \cite{A2} and between the Fibonacci quasicrystal and the Harper model were shown \cite{16}. Recently, the parameter range for a off-diagonal gapless AA model that cannot be mapped onto a QH system was explored \cite{AAH}.  \\
The periodic table of topological insulators has been constructed only for Hermitian Hamiltonians. This is because of the fact that Hermiticity was required in quantum mechanics to guarantee the reality of the spectrum. However,
it was shown in 1998 that Hermiticity requirement is replaced by the analogous
condition of $\mathcal{PT}$ symmetry, where $\mathcal{P}$ and $\mathcal{T}$ operators are parity and time reversal operators, respectively \cite{bender2}. In 2010, a non-Hermitian system was experimentally realized in an optical system \cite{deneyy}. It is well known that a $\mathcal{PT}$ symmetric non-Hermitian Hamiltonian admits real spectrum as long as non-Hermitian degree is below than a critical number. Therefore, a natural question arises. Does there exist a topologically nontrivial system described by non-Hermitian Hamiltonian? This problem was considered by some authors \cite{PTop2,PTop3,PTop4,hensch,PTop1,ekl56}. Hu and Hughes \cite{PTop2} and Esaki et al.  \cite{PTop3} studied non-Hermitian generalization of topologically insulating phase at almost the same time. The main finding of these papers is that topological states don't exist in the $\mathcal{PT}$ symmetric region because of the existence of complex energy eigenvalues. In \cite{PTop2}, they considered Dirac-type Hamiltonians and concluded that the appearance of the complex eigenvalues is an indication of the non-existence of the topological insulator phase in non-Hermitian models. They discussed that finding non-Hermitian topological phases without Dirac-type Hamiltonians could be possible. Esaki et al. considered tight binding honeycomb lattice with imaginary onsite potentials and non-Hermitian generalizations of the Luttinger Hamiltonian and Kane-Mele model \cite{PTop3}. They found that  the zero energy edge states are robust against small non-Hermitian perturbation, while these states decay because of the imaginary part of eigenvalues. In 2012, Ghosh studied topological phase in some non-Hermitian system by changing the metric of the Hilbert space i.e. modifying the inner product \cite{PTop4}. Schomerus considered a one dimensional non-Hermitian tight binding lattice with staggered tunneling amplitude and showed that the system admits complex spectrum \cite{hensch}. Another attempt has recently been made to find non-Hermitian Hamiltonian admitting topological insulator phase \cite{PTop1}. Zhu, Lu and Chen specifically considered non-Hermitian Su-Schrieffer-Heeger model with two conjugated imaginary potential located at the edges of the system \cite{PTop1}. They found that zero energy modes are unstable and corresponding energy eigenvalues are not real in the topologically nontrivial region. In all of these papers, no example of topological edge states for a non-Hermitian system with a real spectrum has been found. The existence of topological phase transition for $\mathcal{PT}$ symmetric non-Hermitian systems is still an important question. In the present paper, we consider a non-Hermitian generalization of the off-diagonal Aubry Andre model. In contrast to general belief, we find topological edge states in the $\mathcal{PT}$ symmetric region, i.e. topological edge states with real spectrum. This is the first example in the literature where topological states are compatible with $\mathcal{PT}$ symmetry. 

\section{Model}

We consider tight binding description of a $\mathcal{PT}$ symmetric lattice \cite{ cem,p14,p15,  p20, p26, p29}. Let us begin with the 1D off-diagonal AA model. This is a 1D tight binding model with modulated tunneling parameter. In addition, suppose two non-Hermitian impurities are inserted in the system. The impurities are balanced gain and loss materials. We require that the non-Hermitian impurities are arranged at symmetrical sites with respect to the center of the lattice, i.e., particles are injected on the $\ds{j}$-th site and removed from the $\ds{(N-j+1)}$-th site, where $\ds{N}$ is the number of lattice sites. The Hamiltonian reads
\begin{eqnarray}\label{mcabjs4}
H=-t\sum_{n=1}^{N-1}\left(1+\lambda \cos{(2\pi \beta n+\Phi)}\right)  a^{\dagger}_{n} a_{n+1}+h.c.\nonumber\\
+i\gamma( a^{\dagger}_j a_j- a^{\dagger}_{N-j+1} a_{N-j+1})
\end{eqnarray}
where the parameter $\gamma$ represents non-Hermitian degree, $\ds{t}$ is unmodulated tunneling amplitude, $\ds{a^{\dagger}_n}$ and $\ds{a_n}$ denote the creation and annihilation operators of fermionic particles on site $\ds{n}$, respectively. The first term in the Hamiltonian is the kinetic energy from the nearest-neighbor tunneling and the last term is the non-Hermitian potential due to two non-Hermitian impurities. The constant $\lambda$ is the strength of the modulation and $\beta$ controls the periodicity of the modulation. The modulation is periodical if $\beta$ is rational and quasi-periodical if it is irrational. As it is usual on the topic of topological insulator in 1D, the modulation phase $\Phi$ is
an additional degree of freedom. \\
Below, we will find energy spectrum of this Hamiltonian numerically. We adopt open boundary conditions with $n = 1$ and $n = N$ being the two edge sites. Note that the translational invariance of the system is broken for open boundary conditions. Before going into the details, let us study the $\mathcal{PT}$ symmetry of the Hamiltonian qualitatively. It is clear that the system has no parity symmetry when the tunneling is modulated quasi-periodically. Therefore, the Hamiltonian is not $\mathcal{PT}$ symmetric when $\beta$ is irrational. Consider now periodical modulation of tunneling. For finite number of $N$, the particles on the two ends of the chain may not have the same tunneling to their nearest neighbors. Furthermore, particles injected on site $j$ and to be lost from site $N-j+1$ may feel different tunneling depending on $\beta$ and $N$. Therefore, we say that the system with edges doesn't generally possess global $\mathcal{PT}$ symmetry in the usual sense. For example, for $\ds{\beta=1/3}$, the system is invariant under usual $\mathcal{PT}$ operation when $N=6$ but not when $N=7$ and $N=8$. It is well known in the theory of $\mathcal{PT}$ symmetric quantum theory that a commuting anti-linear operator can be found for a non-Hermitian  Hamiltonian if the spectrum is real \cite{bendix,bender}. The associated anti-linear symmetry can be interpreted as $\mathcal{PT}$ symmetry. We will show that our system admits real spectrum for some particular cases.\\
\begin{figure}[b]\label{figcem}
{\includegraphics[width=4.5cm]{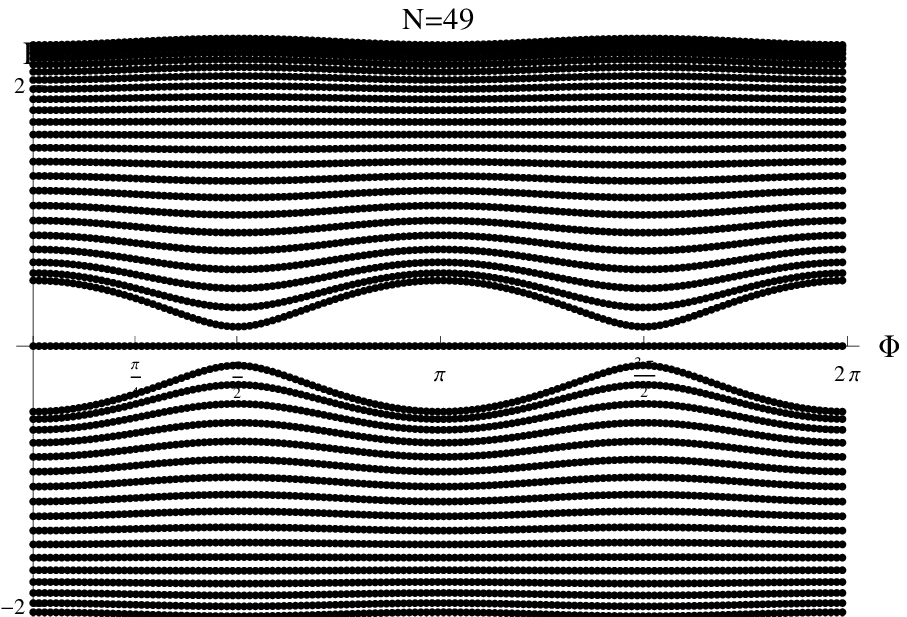},\includegraphics[width=4.5cm]{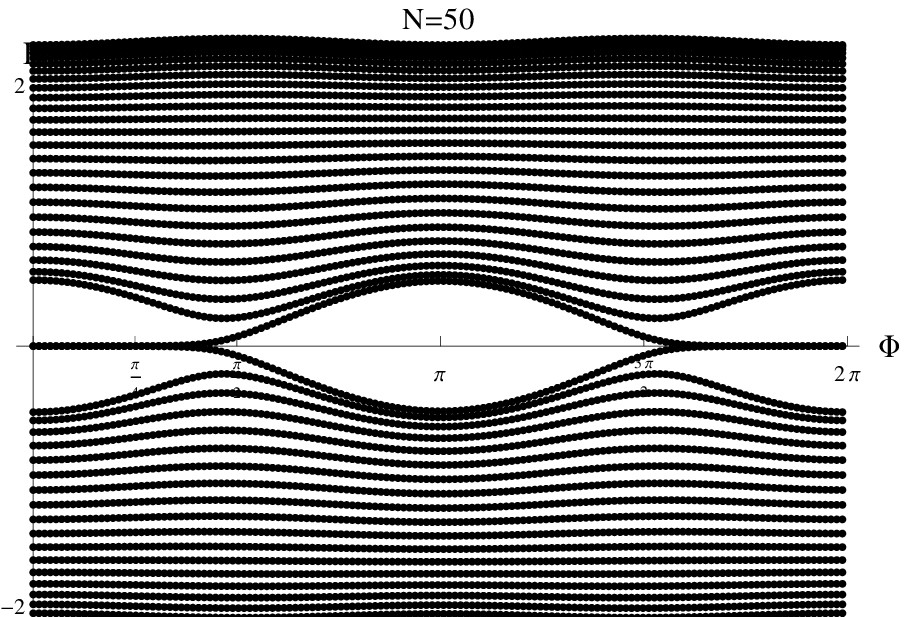}}
\caption{Energy spectrum for $\ds{\lambda=0.4}$, $\ds{\beta=1/2}$ and $\gamma=0$ when $N=49$ and $N=50$ sites. Topological zero energy edge states appear in the whole region of $\Phi$ when the site number is odd. }
\end{figure}
\begin{figure}[t]\label{fig2}
\includegraphics[width=4.3cm]{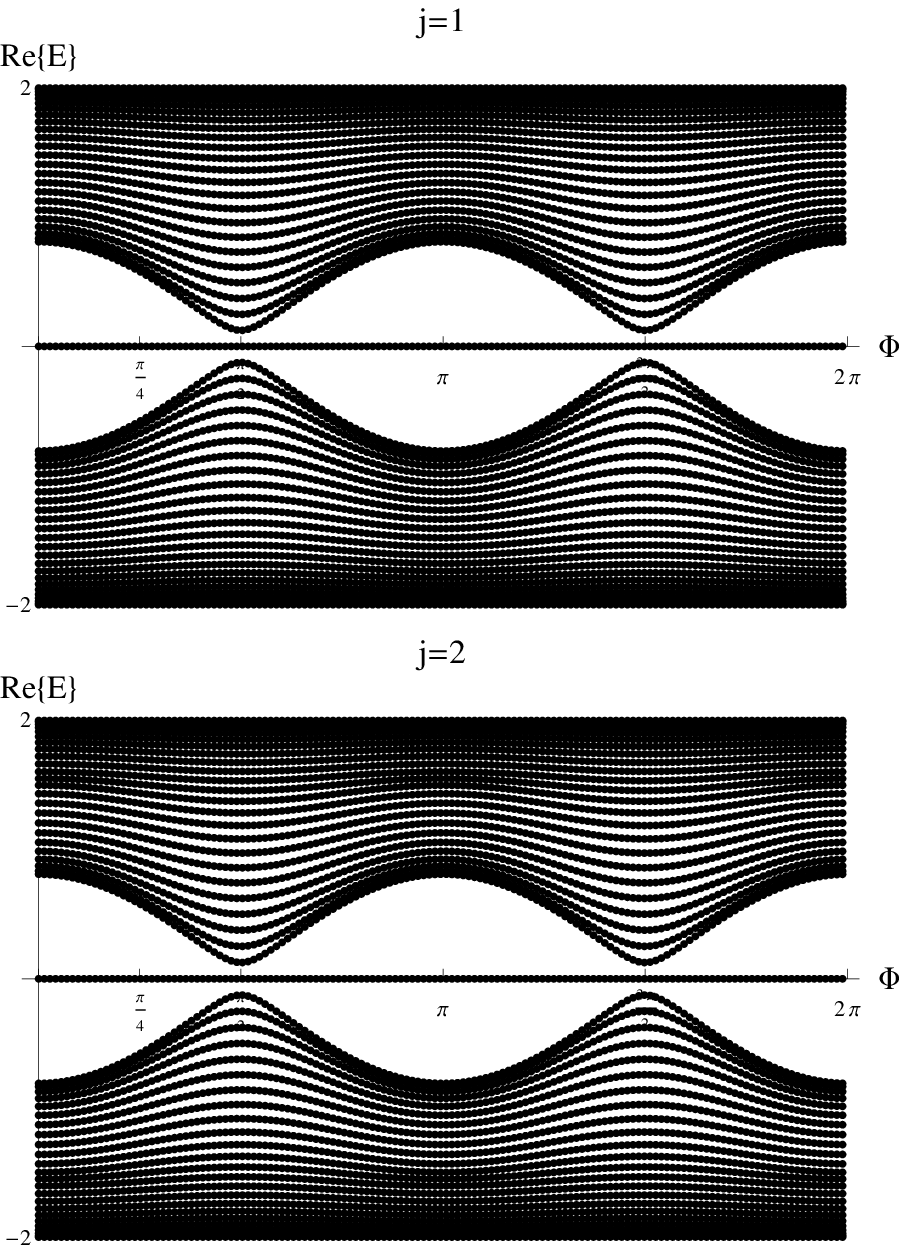},\includegraphics[width=4.4cm]{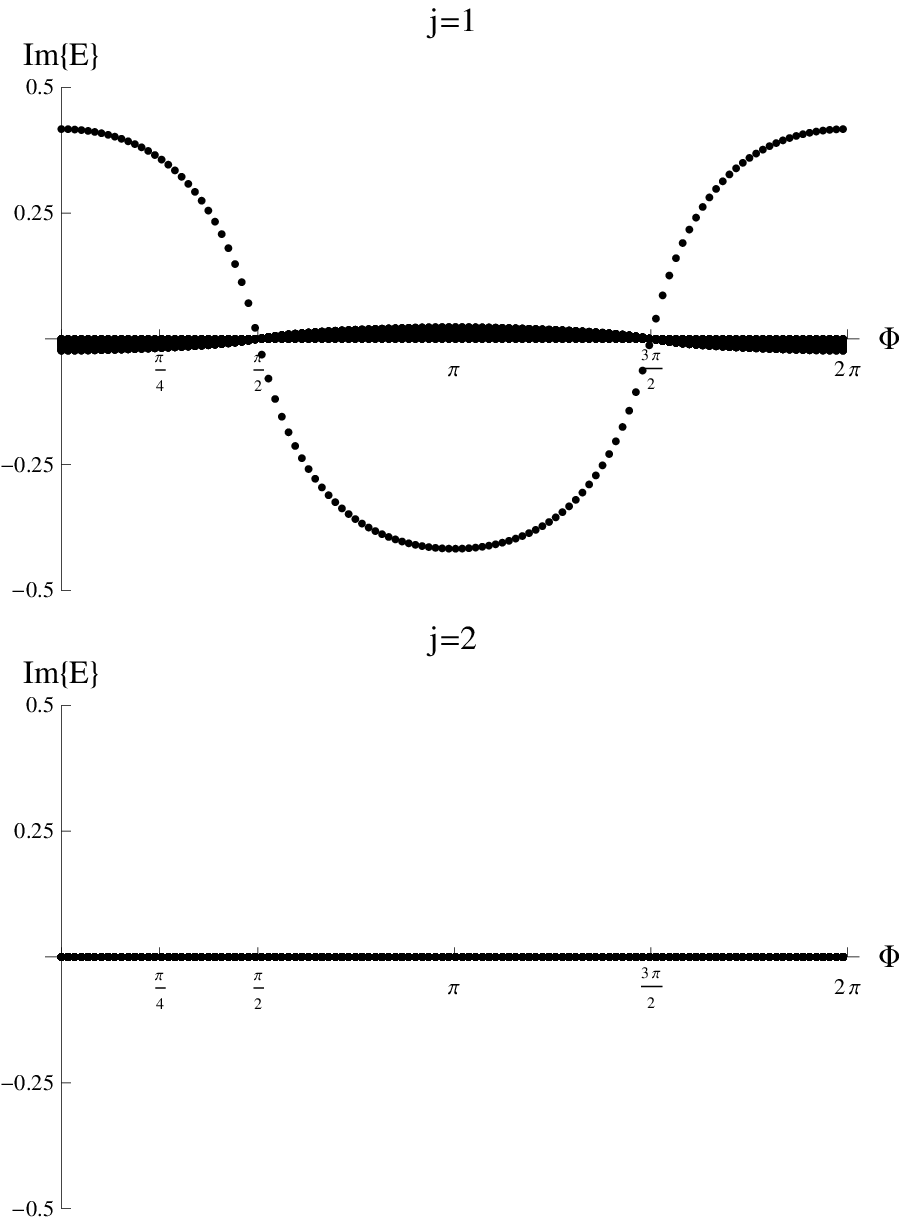},
\includegraphics[width=4.3cm]{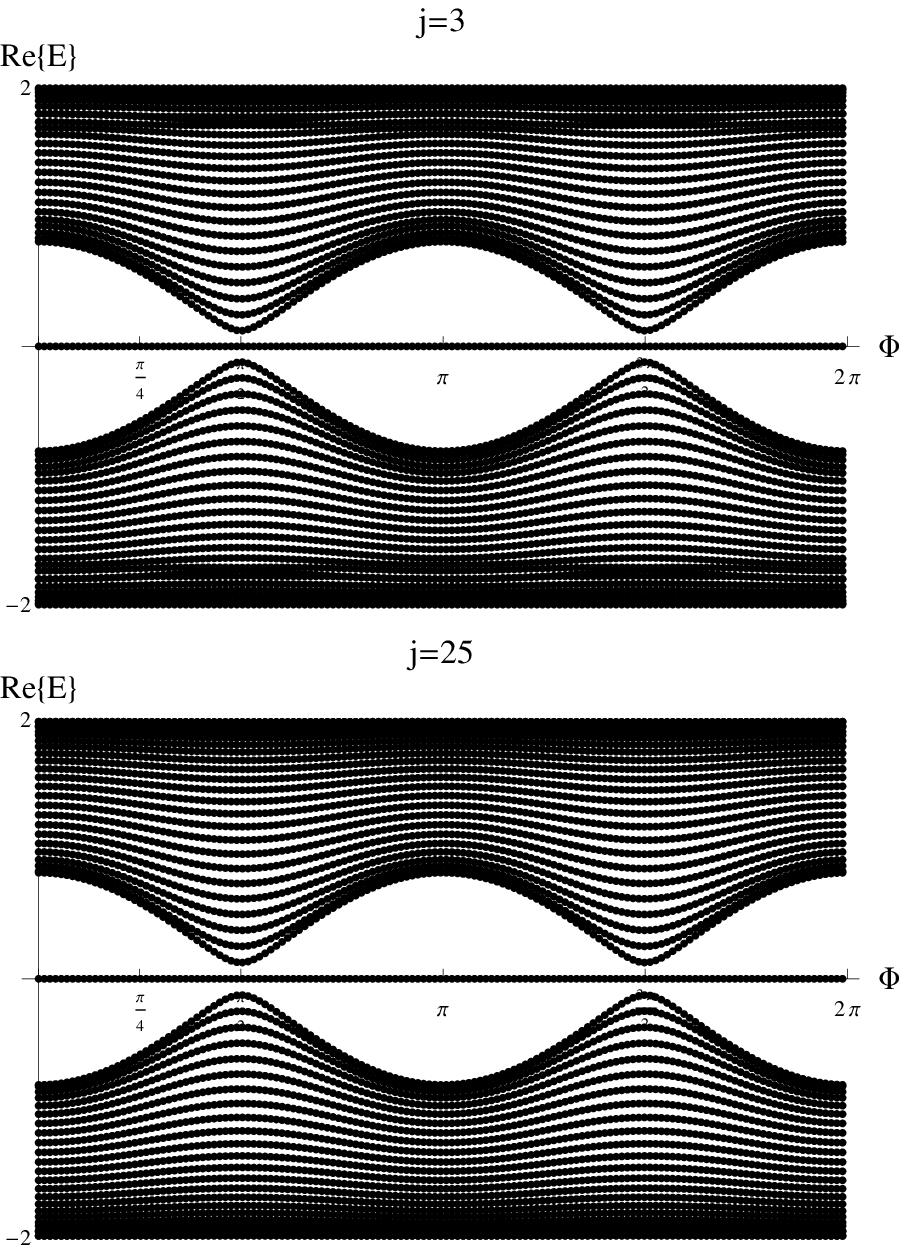},\includegraphics[width=4.4cm]{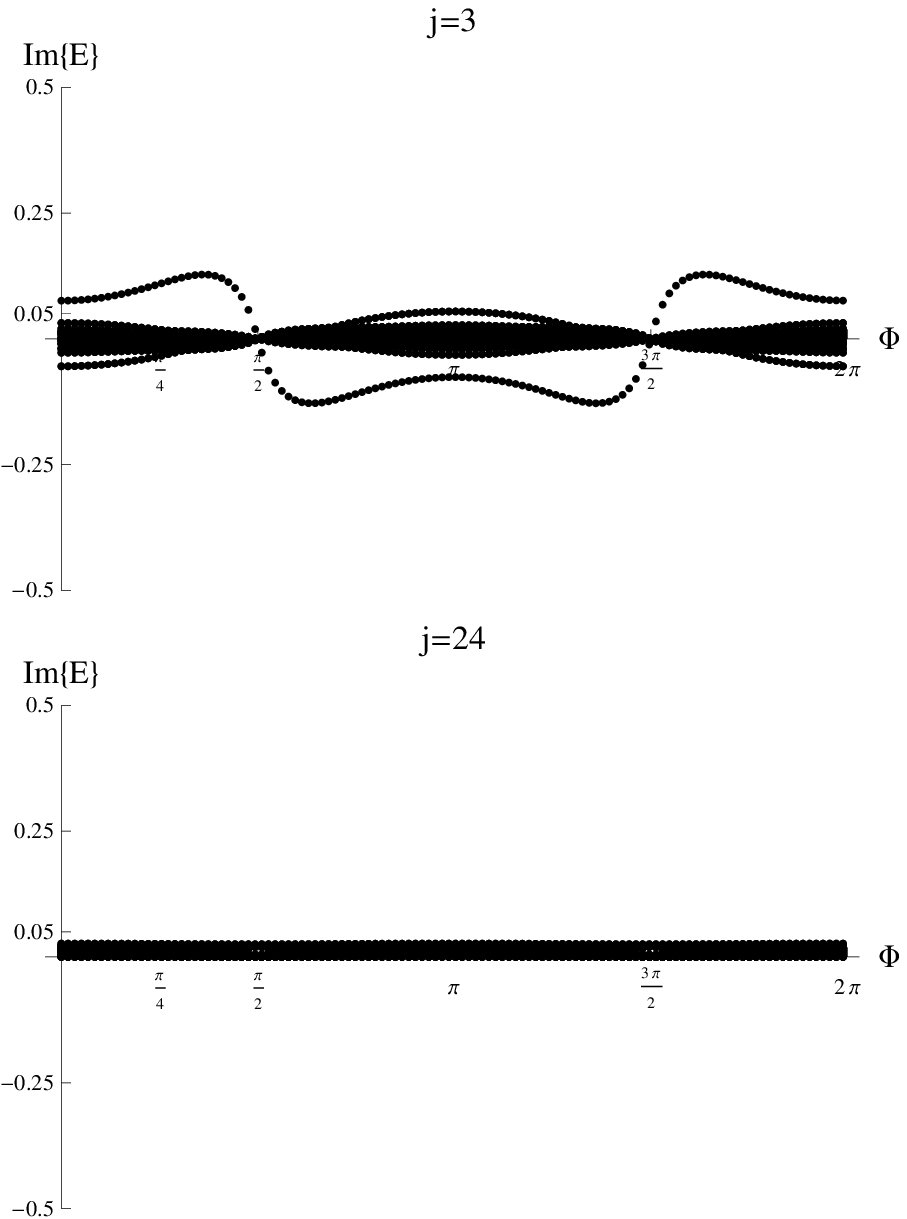}
\caption{The real and imaginary parts of energy spectrum for the parameters $\ds{\lambda=0.4}$, $\ds{\beta=1/2}$, $N=49$ sites and $\ds{\gamma=0.5}$. Observe that topological zero energy states appear in the unbroken $PT$ symmetric phase if $\ds{j=2}$. Note that the spectrum would be real valued when $j=24$ if $\ds{\gamma <  0.423}$.}
\end{figure}
\begin{figure}[t]
\label{fig1}
\includegraphics[width=4.3cm]{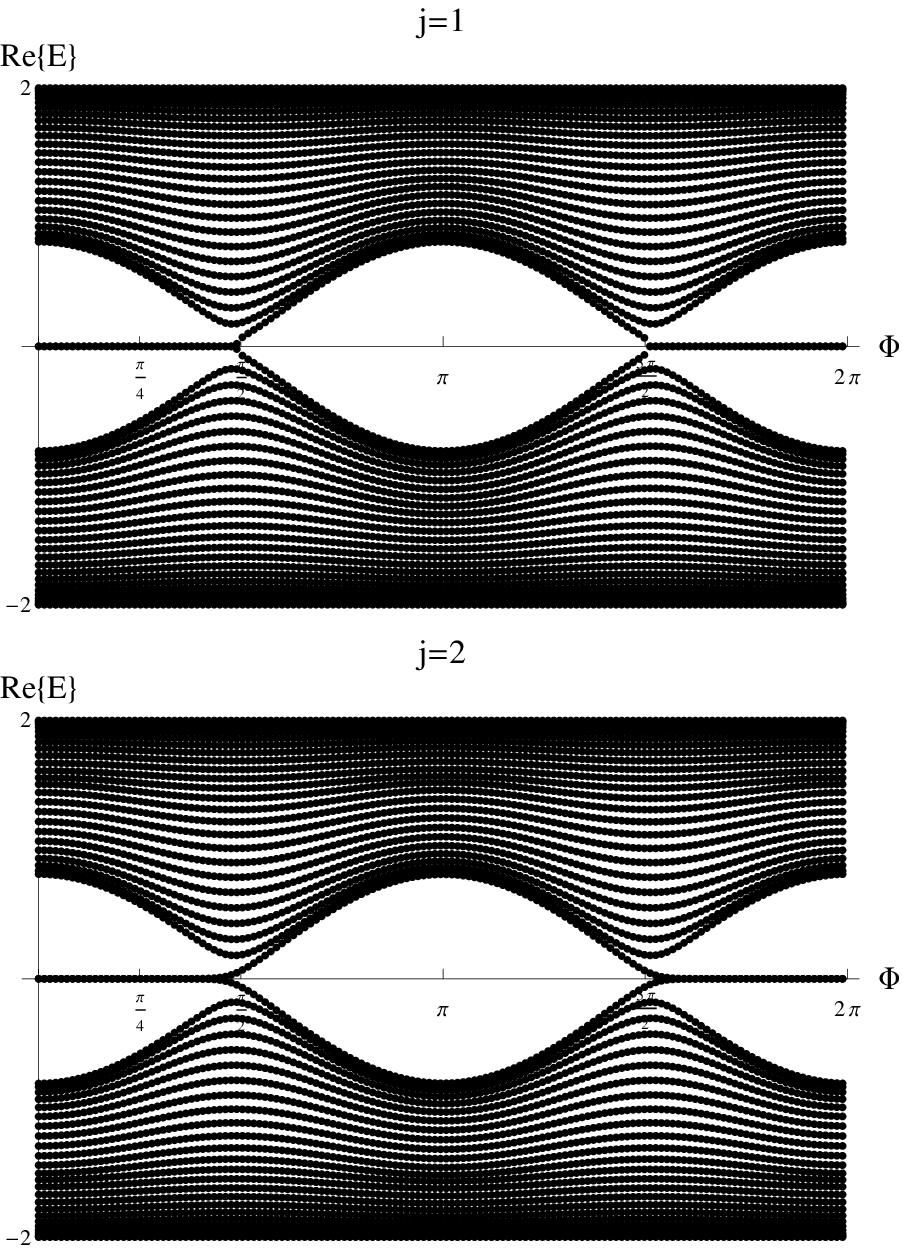},\includegraphics[width=4.4cm]{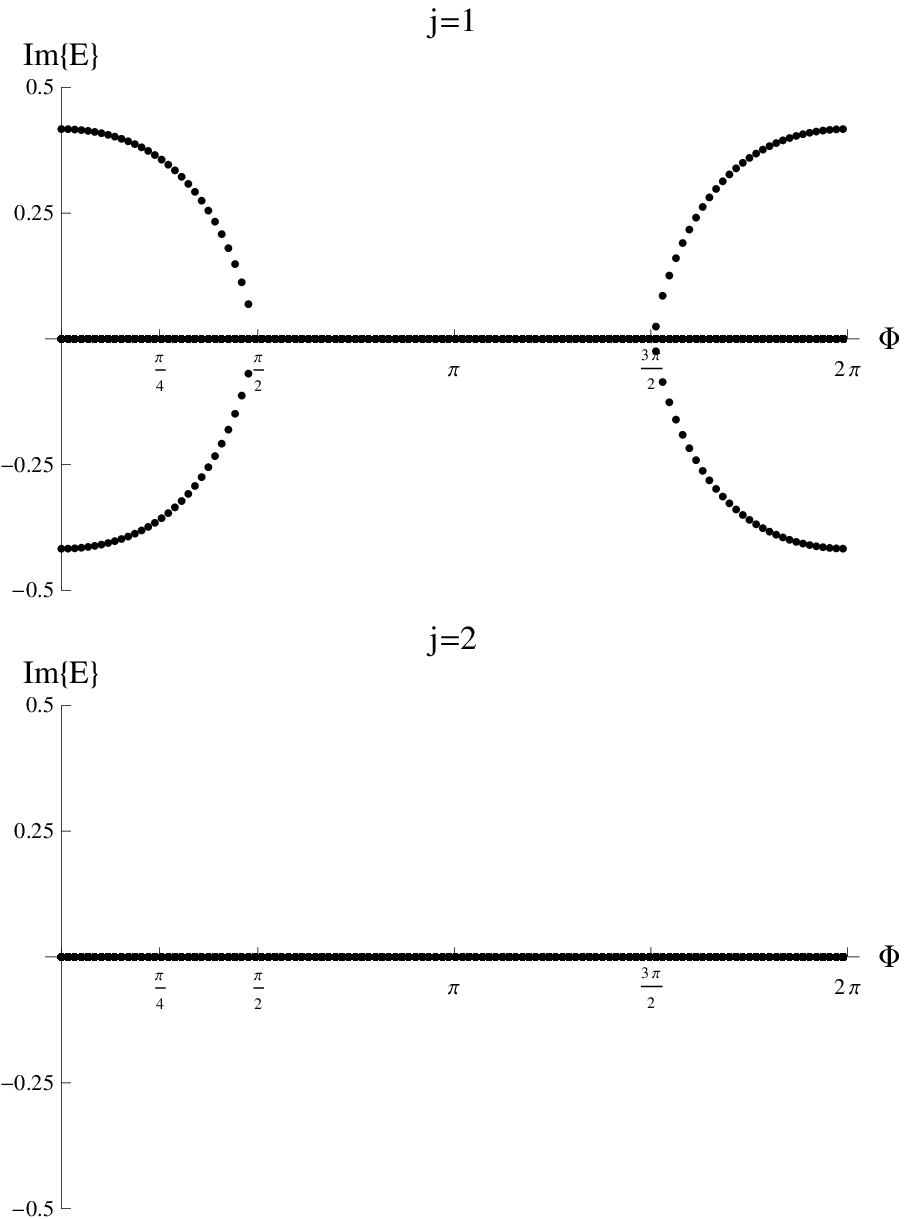},
\includegraphics[width=4.3cm]{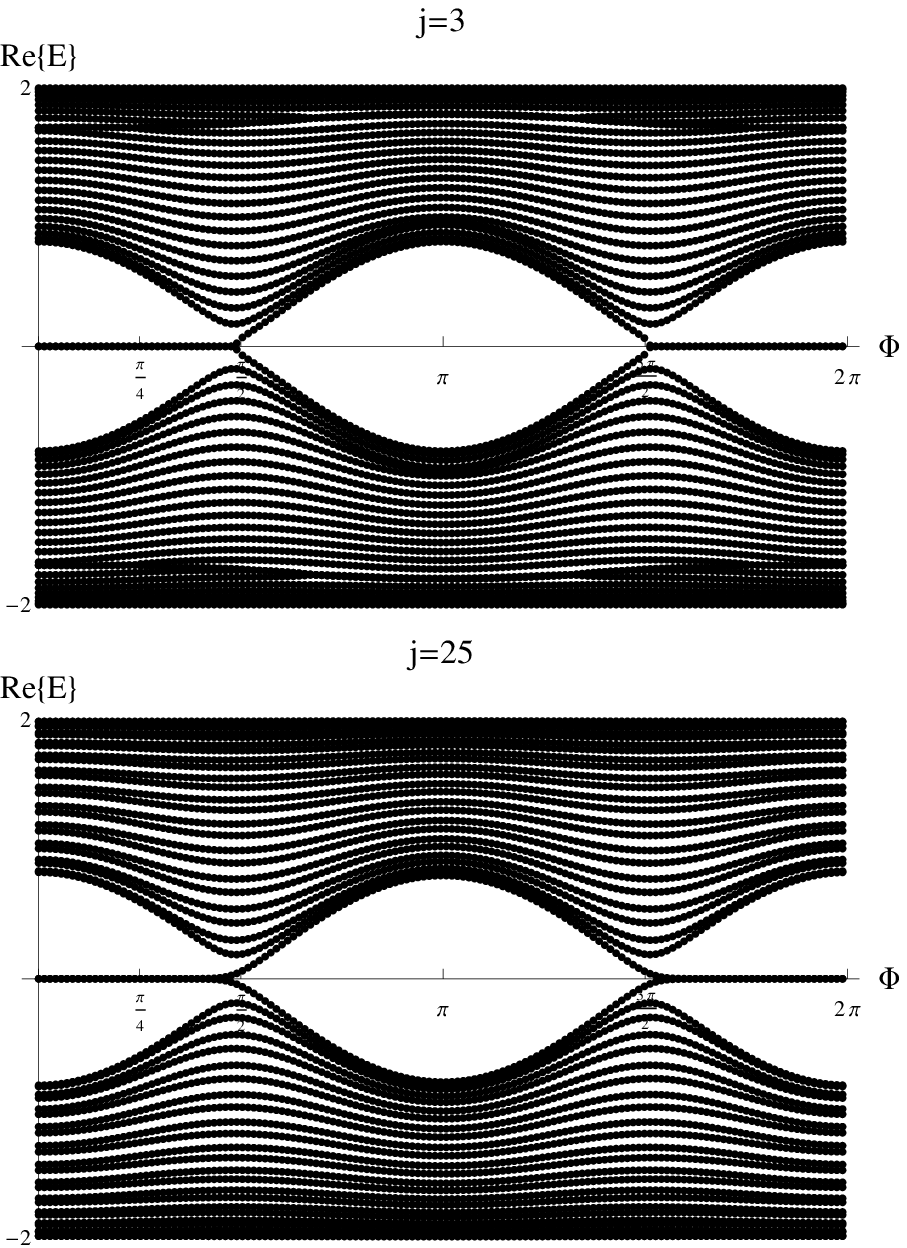},\includegraphics[width=4.4cm]{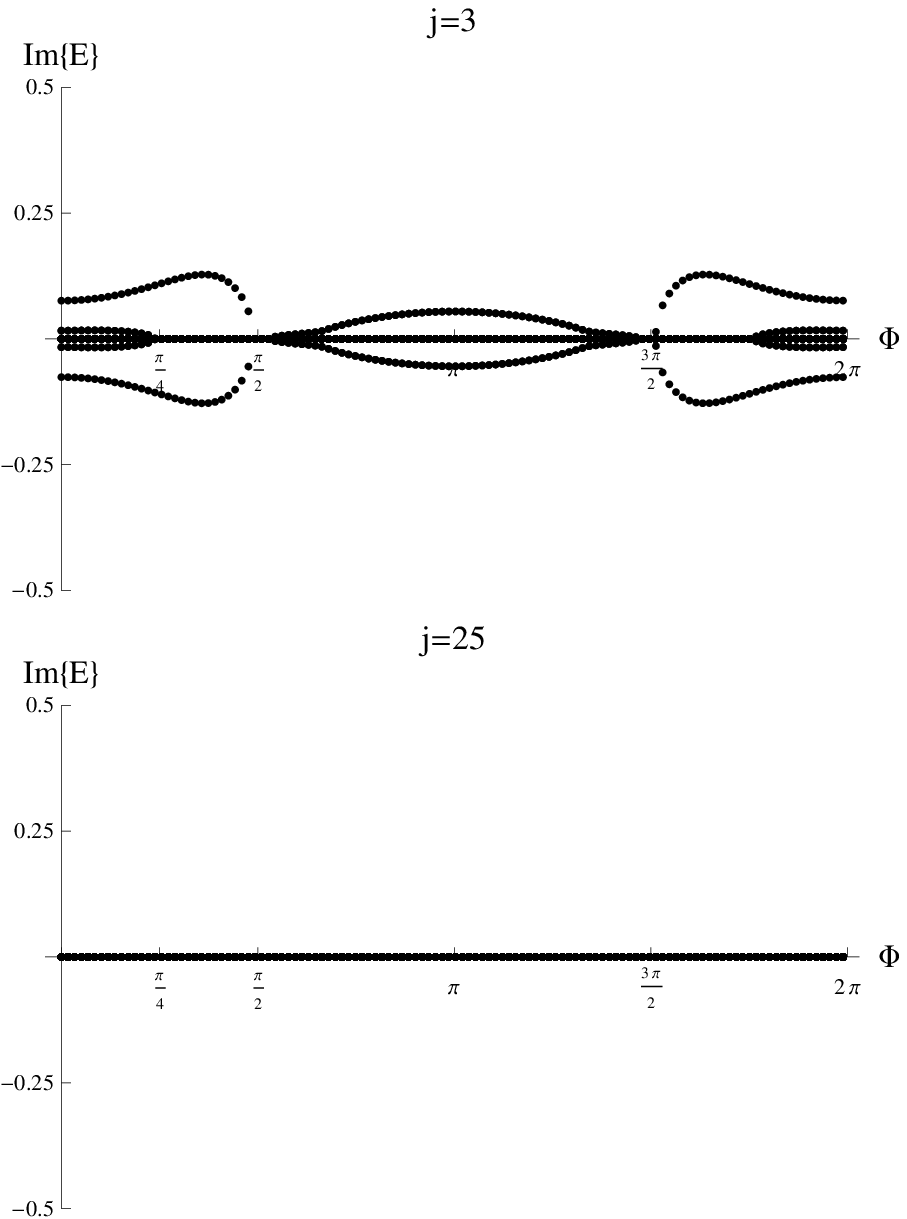}
\caption{The real and imaginary parts of energy spectrum for the parameters $\ds{\lambda=0.4}$, $\ds{\beta=1/2}$, $N=50$ sites and $\ds{\gamma=0.5}$. At this value of the non-Hermitian degree, topological zero energy edge states appear in the unbroken $PT$ symmetric phase only if the non-Hermitian impurities are placed either on the neighbors of the edges, i.e., $j=2$ or the neighbors of the center of the lattice, i.e. $j=25$. If $j=1$ or $j=3$, the spectrum is not real anymore. Contrary to the case with $j=1$, the spectrum is complex in both topologically trivial and nontrivial regions when $j=3$.} 
\end{figure}
Our aim is to look for a topologically nontrivial phase with real spectrum for our non-Hermitian Hamiltonian. Since the $\mathcal{PT}$ symmetry is broken for quasi-periodical tunneling, we consider periodical tunneling parameter. Let us start with $\ds{\beta=1/2}$, where the tunneling alternates at fixed $\Phi$ between the two values $\ds{1+\lambda \cos(\Phi)}$ and $\ds{1-\lambda \cos(\Phi)}$. Hence, the corresponding spectrum has two energy bands. Suppose first that the system is Hermitian, $\ds{\gamma=0}$. In this case, our model is reduced to the well known  Su-Schreier-Heeger model (SSH)  \cite{SSH}. We plot the energy spectrum as a function of the phase $\Phi$ in the Fig.1. As can be seen, the system has maximum band gap at $\Phi=0$ and the bulk gap closes and reopens as $\Phi$ changes. Remarkably, zero energy states appear in the spectrum. It is well known that these zero-energy states are localized around the two edges of the system \cite{AAH}. This is a signature of topologically nontrivial phase. The region of nontrivial topological phase depends on whether the number of lattice sites is odd or even. If $\ds{N}$ is odd, zero energy edge states appear for all $\ds{\Phi}$. However, this is not case if $\ds{N}$ is even. The zero energy edge state exists at $\ds{\Phi=0}$ and varying $\ds{\Phi}$ leads to the topological phase transitions that occur at $\ds{\Phi=\pi/2}$ and $\ds{\Phi=3\pi/2}$. The region $\ds{\pi/2<\Phi<3\pi/2}$ corresponds to topologically trivial phase. Let us now study the existence of edge states if the two non-Hermitian impurities are introduced into the system. We are looking for topological edge states with real energy eigenvalues. We expect that the effect of impurities on the real part of energy spectrum is not appreciable provided $N$ is large and $\gamma$ is not much bigger than the tunneling parameter $t$. To check this idea and study the reality of the spectrum, we present numerical calculation for odd and even number of lattice sites. We see that zero energy states exist even in the presence of the non-Hermitian impurities. However, we are not interested in zero energy states in broken $\mathcal{PT}$ symmetric region. We are looking for such states in the unbroken $\mathcal{PT}$ symmetric region. The key point here is the site position of particle injection and removal. More precisely, where we inject and remove particles has almost nothing to do with the real part of spectrum but plays a vital role on the reality of the spectrum. This can be seen clearly in the Fig.2 $\ds{(N=49)}$ and Fig.3 $\ds{(N=50)}$. Let us start our discussion with the case of $\ds{j=1}$, i.e., the particles are injected to the one edge and lost from the other edge. In this case, the $\mathcal{PT}$ symmetric phase is spontaneously broken and the spectrum becomes complex in the topologically nontrivial region whereas the spectrum is real as long as $\gamma$ is below than a critical number in the topologically trivial region. As a result, we say that the spectrum is complex in the whole region of the phase $\ds{0<\Phi<2\pi}$ (in the  region $\ds{\pi/2<\Phi<3\pi/2}$) when $N$ is an odd (even) number. Secondly, let us now suppose that the gain/loss occurs at $\ds{j=j^{\prime}=m/\beta}$, where $\ds{m=1,2,..}$ are positive integers ($\ds{j^{\prime}=2,4,6,..}$ for $\ds{\beta=1/2}$). It is remarkable that the spectrum becomes real in the whole region of the phase provided that non-Hermitian degree $\gamma$ is below than a critical number, $\ds{\gamma_c}$. The critical numbers for even and odd values of $N$ are almost the same and slightly changes if we increase $N$. However, it decreases with the modulation strength $\ds{\lambda}$ and vanishes if $\ds{\lambda>1}$. At fixed $\ds{N}$ and $\ds{\lambda}$, $\gamma_c$ decreases also with increasing $m$. The maximum value of $\ds{\gamma_c}$ occurs when the non-Hermitian impurities are on the nearest neighbors of the edges, i.e., at $j=2$. ($\ds{\gamma_c=0.56}$ for $\lambda=0.4$ and decreases with increasing $\lambda$). As a final step, we numerically confirm that the zero energy states are localized around the two edges of the lattice. The main finding of this paper is that our system has nontrivial topological edge states in the unbroken $\mathcal{PT}$ symmetric region. To this end, we note that $\ds{\gamma_c}$ is either very small or zero when $\ds{j}$ is not equal to $\ds{j^{\prime}}$. The only exception is the one when the impurities are on the two symmetrical neighboring sites with respect to the center of the lattice. In this case, $\ds{\gamma_c=0.43}$ when $N=49$ and $\ds{\gamma_c=0.6}$ when $N=50$. If we change $N$, the critical value $\gamma_c$ changes drastically.\\
We show that the position of non-Hermitian impurities has dramatic effect on the reality of the spectrum for $\ds{\beta=1/2}$. We can extend our previous analysis for other values of $\ds{\beta=1/p}$, where $p>2$ is a positive integer. In this case, energy spectrum is split into $p$ bands. We stress that topological zero energy edge modes don't exist when $p$ is an odd number. Below we first study the reality of the spectrum when $p$ is an odd number and then discuss topological phase when $p$ is an even number. As an example consider $\ds{p=3}$, where there are three bands that are crossed as the phase $\Phi$ sweeps from zero to $2\pi$. In the presence of impurities, the real part of the spectrum is almost left unchanged provided $\gamma$ is not much bigger than tunneling amplitude. Compared to the case with $\beta=1/2$, the impurity lattice sites necessary for the reality of the spectrum are shifted according to the formula $\ds{j=j^{\prime}=m/\beta=3,6,9,...}$, where the critical number $\gamma_c$ decreases with increasing $\ds{j^{\prime}}$. We note that the site number $N$ plays also a role on the reality of the spectrum. To understand better, suppose that particles on the one edge of the system feel tunneling with amplitude $\ds{1+\lambda \cos(\Phi)}$. Depending on the number $N$, particles on the other edge of the lattice feel either the same tunneling or one of following tunnelings with amplitudes $\ds{1+\lambda \cos(2\pi/3+\Phi)}$, $\ds{1+\lambda \cos(\pi/3+\Phi)}$. So, there are three different configurations at fixed $j^{\prime}$ and we should calculate the spectrum separately for three different consecutive numbers. As an example, we take $N=48$, $N=49$ and $N=50$ and $j=3$. We numerically see that the spectrum is real when $N=50$ unless $\gamma>\gamma_c=0.45$ (at $\lambda=0.4$) and $\mathcal{PT}$ symmetry is broken since the spectrum is complex for $N=48$ and $N=49$. Analogously, we expect a similar picture if $\ds{N \rightarrow N\mp3}$ since the same configuration is obtained. Therefore, the spectrum is real if N equals to .$..,47,50,53,..$. and $\gamma<\gamma_c$, where the critical number $\gamma_c$ slightly changes with $N$. Suppose now that $p$ is an even number. We are mainly interested in this case since topological zero energy states are available. Let us study the reality of the spectrum. As an example, consider $p=4$, where the impurity lattice sites necessary for the reality of the spectrum are given by $\ds{j^{\prime}=4,8,..}$. The spectrum becomes real if $N=. ..,51,55,59, ..$. and the critical values $\ds{\gamma_c}$ changes slightly with $N$ at fixed $\lambda$. The critical value $\ds{\gamma_c\approx 0.15}$ for $p=4$ (at $\lambda=0.4$ and $j^{\prime}=4$ ) and decreases with $p$. As a result, we say that topological zero energy states exist when $p$ is even and we show that the corresponding spectrum becomes real under certain conditions.\\
It is worthwhile to point out that the topological phase in the unbroken $\mathcal{PT}$ symmetric phase can be broken by the next-nearest-neighbor tunneling. We introduce next-nearest-neighbor tunneling to the off-diagonal $\mathcal{PT}$ symmetric AA model
model. The new Hamiltonian reads
\begin{eqnarray}\label{mcabjs4ek}
H^{\prime}=H+t^{\prime}\sum_{n=1}^{N-2}  a^{\dagger}_{n} a_{n+2}+h.c.
\end{eqnarray}
where $H$ is given by (\ref{mcabjs4}) and $\ds{t^{\prime}}$ is the next-nearest-neighbor tunneling constant. Suppose the next-nearest-neighbor tunneling parameter is site independent. Without any loss of generality we assume the next-nearest-neighbor tunneling is constant. The energy of the edge states are not zero anymore and the particle-hole symmetry is lost since next-nearest-neighbor tunneling contributes the energy spectrum perturbatively. Introducing non-Hermitian impurities into the system breaks the $\mathcal{PT}$ symmetry spontaneously in the presence of next-nearest-neighbor tunneling. Although edge states are still robust as a direct
result of the topological nature, complex eigenvalues appear in the system. Fortunately, the imaginary part of energy eigenvalues are very small in a broad range of $\gamma$ if $\ds{j=j^{\prime}}$. For example, for the parameters $t=1$, $t^{\prime}=0.1$, $\ds{\lambda=0.4}$, $\ds{\beta=1/2}$, $j=2$ and $N=50$, the imaginary part of energy eigenvalues are at the order of $\ds{10^{-4}}$ when $\ds{\gamma=0.2}$ and $\ds{10^{-3}}$ when $\ds{\gamma=0.3}$.\\
\begin{figure}[t]
\label{fig000}
{\includegraphics[width=4.5cm]{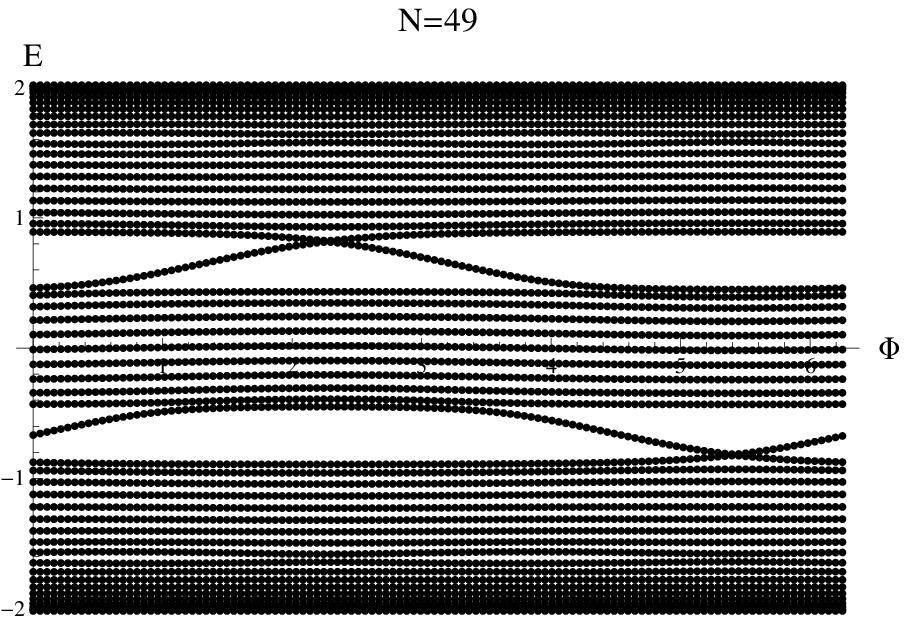},\includegraphics[width=4.5cm]{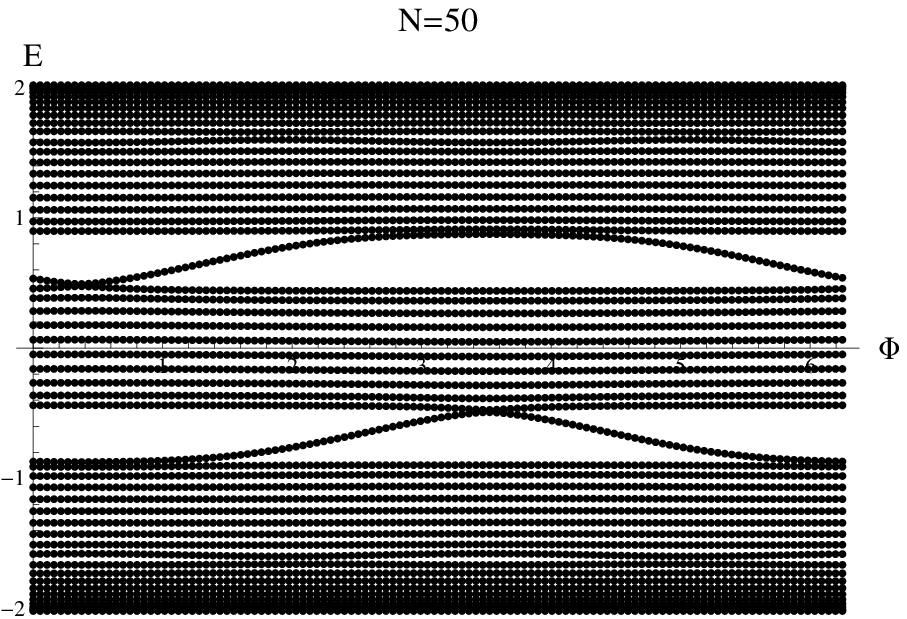}}
\caption{The spectrum for the diagonal AA model for $t=1$, $\ds{V=0.4}$ $\ds{\beta=\sqrt{13}-3}$. The site number $N$ changes the particular values of the phase $\Phi$ where the bands are crossed.}
\end{figure}
So far, we have omitted quasi-periodical modulation of the tunneling parameter since the system is not $\mathcal{PT}$ symmetric when $\beta$ is irrational. Here, we prove numerically that the spectrum is complex for this case. In the absence of non-Hermitian impurities, the energy spectrum is broken into fractal set of bands and gaps due to the quasi periodical nature of the tunneling parameter. As shown in the fig.4, the bands remains almost the same but they are connected by the edge states as $\Phi$ shifts adiabatically from zero to $\ds{2\pi}$. The states within the band gap are localized states  either on the right or the left edge of the lattice, while the states within the bands are extended. In the presence of non-Hermitian potential, we numerically check that the real part of the spectrum is almost unchanged as long as the non-Hermitian degree is weak. However, we see that complex eigenvalues appear whenever $\gamma$  is different from zero. We repeat our calculations for different values of site number $\ds{j}$ and we see no signature of nontrivial topological phase with real spectrum. \\
As an another example where quasi-periodicity breaks the $\mathcal{PT}$ symmetry, let us consider a non-Hermitian generalization of the diagonal Aubry Andre model
\begin{eqnarray}\label{mcabsf}
H= -t \sum_{n=1}^{N-1}a^{\dagger}_{n} a_{n+1}+h.c.+\sum_{n=1}^NV \cos{(2\pi \beta n+\Phi)}a^{\dagger}_{n} a_{n}\nonumber\\
 +i\gamma( a^{\dagger}_j a_j- a^{\dagger}_{N-j+1} a_{N-j+1})
\end{eqnarray}
where we suppose $\beta$ is irrational. For this Hamiltonian, tunneling parameter is constant but the onsite potential is quasi-periodic. It is well known that the Hermitian
model exhibits a localization transition at $V=2t$ when the modulation is quasi-periodic. All bulk eigenstates
are extended for $\ds{0 < V < 2t}$ and localized for
$V > 2t$. The energy spectrum as a function of $\Phi$ consists of fractal set of bands and within the band gap edge states appear. In the presence of non-Hermitian impurities, the spectrum becomes complex. We perform extensive numerical calculation but we did not find topologically nontrivial phase with real spectrum.\\
We have discussed that topological zero energy edge states appear in our system when $\ds{\beta=1/2,1/4,..}$. To understand the topological origin of the edge states for $\ds{\beta=1/2}$, let us rewrite the Hamiltonian (1) in the Majorana basis, where the annihilation operators at site $2n$ and $2n+1$ are defined by $\ds{a_{2n}=\sigma_{2n}+i\tau_{2n}}$, $\ds{a_{2n+1}=\tau_{2n+1}+i\sigma_{2n+1}}$ \cite{AAH}. The two new operators $\sigma_j$ and $\tau_j$ are two species of Majorana fermions at site $j$. Therefore the Hamiltonian (1) without the non-Hermitian term becomes $H_0=\sum_n\Delta_{+} (\sigma_{2n}\sigma_{2n+1} -\tau_{2n}\tau_{2n+1}) +\Delta_{-} (\sigma_{2n}\sigma_{2n-1} -\tau_{2n}\tau_{2n-1}  ) $, where $\ds{\Delta_{\mp}=-2it(1\mp\cos\Phi)}$. This Hamiltonian describes two uncoupled identical Majorana chains and the system has $Z_2$ topological index when $\ds{\Delta+>\Delta_-}$. The Majorana physics is most easily understood in the limit $\ds{\Delta_-=0}$. In this case, the system has two unpaired Majorana operators in each uncoupled chains, $\ds{\sigma_1}$,  $\ds{\sigma_N}$ and  $\ds{\tau_1}$,  $\ds{\tau_N}$. So, zero energy Majorana fermions localized at the edges appear. However, non-Hermitian interaction term includes a coupling term between these unpaired Majorana operators if non-Hermitian impurities are placed at the edges, $j=1$. Therefore the corresponding eigenvalues become complex valued. If the non-Hermitian impurities are placed such that $j=2$, then the non-Hermitian term of the Hamiltonian (1) in the Majorana basis contains a coupling term $\sigma_{2}\tau_{2} -\sigma_{N-1}\tau_{N-1}$. Since it doesn't contain unpaired Majorana operators, the system admits real energy eigenvalues if the non-Hermitian degree is smaller than a critical value. \\
To sum up, we have studied the non-Hermitian generalization of Aubry Andre model. In this paper, we have shown a novel topological aspect for the non-Hermitian generalization of the Aubry Andre model. We have shown that the system can undergo a topological phase transition in the $\mathcal{PT}$ symmetric region by varying the lattice position of the non-Hermitian impurities.  We think photonic crystals could be used to observe the topological edge states for our model. Our system is the first example in the literature and it is worth to look for other topologically nontrivial $\mathcal{PT}$ symmetric systems with real spectra. Another important task is to extend the periodic table of topological insulator to non-Hermitian Hamiltonians.

\end{document}